\documentclass[aps,prl,reprint,superscriptaddress,amsmath,amssymb,floatfix,longbibliography]{revtex4-2}
\usepackage[T1]{fontenc}
\usepackage[utf8]{inputenc}
\usepackage{graphicx}
\usepackage{bm}
\usepackage{microtype}
\usepackage[hidelinks,unicode]{hyperref}
\graphicspath{{figures/}}
\begin{document}
\title{Moir\'e Excitons in a Chern Ferromagnet}
\author{Shiyu Peng}
\thanks{These authors contributed equally.}
\affiliation{Department of Physics, The Chinese University of Hong Kong, Hong Kong, China}
\affiliation{State Key Laboratory of Quantum Information Technologies and Materials, The Chinese University of Hong Kong, Hong Kong, China}
\author{Yifan Wang}
\thanks{These authors contributed equally.}
\affiliation{Department of Physics, The Chinese University of Hong Kong, Hong Kong, China}
\affiliation{State Key Laboratory of Quantum Information Technologies and Materials, The Chinese University of Hong Kong, Hong Kong, China}
\author{Peiyao Qin}
\affiliation{Department of Physics, The Chinese University of Hong Kong, Hong Kong, China}
\affiliation{State Key Laboratory of Quantum Information Technologies and Materials, The Chinese University of Hong Kong, Hong Kong, China}
\author{Yixuan Liu}
\affiliation{Department of Physics, The Chinese University of Hong Kong, Hong Kong, China}
\affiliation{State Key Laboratory of Quantum Information Technologies and Materials, The Chinese University of Hong Kong, Hong Kong, China}
\author{Xi Dai}
\affiliation{Department of Physics, The Hong Kong University of Science and Technology, Clear Water Bay, Hong Kong, China}
\author{Chen Hu}
\email{chenhu@cuhk.edu.hk}
\affiliation{Department of Physics, The Chinese University of Hong Kong, Hong Kong, China}
\affiliation{State Key Laboratory of Quantum Information Technologies and Materials, The Chinese University of Hong Kong, Hong Kong, China}

\begin{abstract}

Recent experiments have established excitonic optical resonances as powerful probes and control knobs of moiré Chern ferromagnets. However, the microscopic nature of excitons formed on such interaction-driven, time-reversal-symmetry-broken ground states remains largely unexplored. In this paper, using a Gutzwiller plus Bethe-Salpeter equation framework, we investigate twisted bilayer MoTe\textsubscript{2} at hole filling \(\nu_{h} = 1\) and identify two near-infrared Chern-ferromagnet excitons (CFEs) selectively coupled to opposite circular polarizations. Strong electron-hole interactions markedly enhance their energy separation and drive a pronounced momentum-space reconstruction of the emergent X\textsuperscript{-} state, beyond the quasiparticle-transition picture. We further find that this reconstruction enhances the interlayer electron-hole character of X\textsuperscript{-}, giving rise to a giant quadratic Stark response. These findings reveal the nontrivial interplay between electron-electron correlations and electron-hole interactions underlying CFE formation, providing optical fingerprints of moiré Chern-ferromagnetic order and a microscopic basis for its optical control.

\end{abstract}

\maketitle


\nocite{ref1,ref2,ref3,ref4,ref5,ref6,ref7,ref8,ref9,ref10,ref11,ref12,ref13,ref14,ref15,ref16,ref17,ref18,ref19,ref20,ref21,ref22,ref23,ref24,ref25,ref26,ref27,ref28,ref29,ref30,ref31,ref32,ref33,ref34,ref35,ref36,ref37,ref38,ref39,ref40,ref41,ref42,ref43,ref44,ref45,ref46}

Moiré superlattices formed by stacking atomically thin crystals with a small twist or lattice mismatch provide a versatile platform for creating emergent electronic states in two dimensions (2D) \cite{ref1}. The resulting long-wavelength moiré potential can strongly reconstruct the low-energy electronic structure, giving rise to narrow bands and a broad range of interaction-driven quantum phenomena. In semiconducting transition-metal dichalcogenide (TMD) bilayers, electron--electron (\emph{e}-\emph{e}) interactions play a central role, producing a rich set of filling-dependent correlated states including Mott insulators, generalized Wigner crystals, and magnetism \cite{ref2,ref3,ref4,ref5,ref6,ref7,ref8,ref9,ref10,ref11}. An especially striking regime arises when strong correlations coexist with nontrivial band topology: narrow valley Chern bands in twisted \(\text{MoTe}_{2}\) host integer and fractional quantum anomalous Hall states \cite{ref11,ref12,ref13,ref14,ref15,ref16,ref17}. At integer hole fillings, strong \emph{e}-\emph{e} interactions can spontaneously break time-reversal symmetry and stabilize spin- and valley-polarized topological ground state, forming moiré Chern ferromagnets.

\begin{figure}[t]
\centering
\includegraphics[width=\columnwidth]{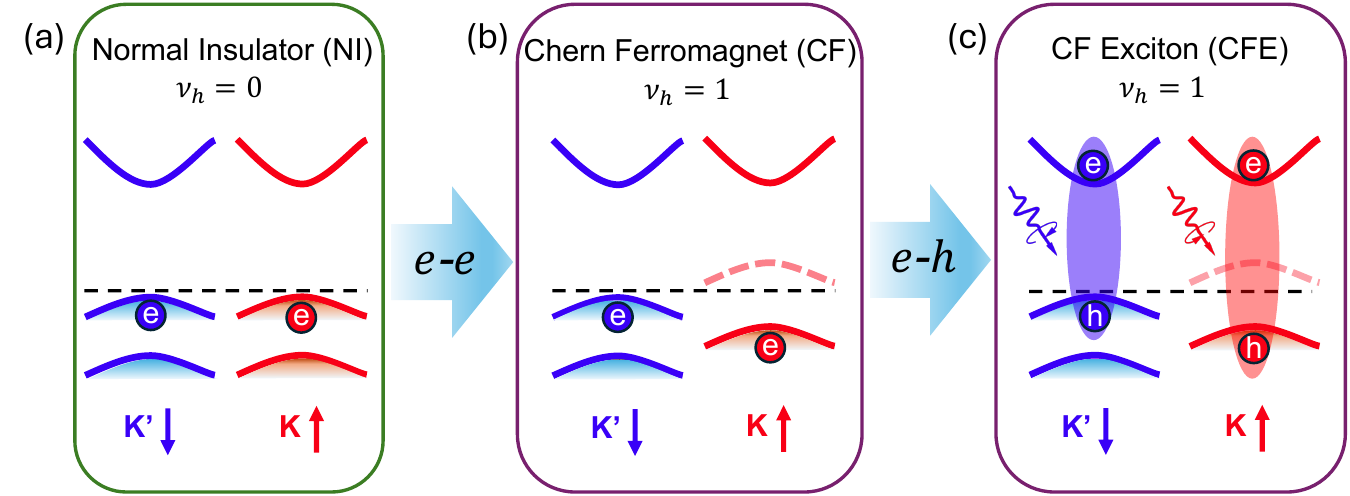}
\caption{Schematic illustration of Chern-ferromagnet excitons (CFEs). Blue and red denote the K' and K valleys, respectively, with opposite spin character. (a) At hole filling \(\nu_{h} = 0\), the undoped system forms a valley-degenerate normal insulator (NI), used here as the reference state for comparison. (b) At \(\nu_{h} = 1\), electron--electron interactions spontaneously break time-reversal symmetry and stabilize a valley-polarized Chern ferromagnet (CF), with the topmost K-valley valence band empty of electrons (hole occupied; dashed red line). (c) Optical excitation of the Chern ferromagnets creates bound electron-hole pairs, i.e. CFEs. The two CFEs originate from opposite valleys and couple selectively to opposite circular polarizations of light. Black dashed lines mark the Fermi level.}
\label{fig:1}
\end{figure}

Optical excitation provides direct access to the excited-state many-body physics of moiré Chern ferromagnets. In 2D TMDs, strong electron-hole (\emph{e-h}) interactions generates tightly bound excitons that govern the optical response \cite{ref18,ref19}. In TMD moiré superlattices, the moiré potential further reconstructs their internal spatial structure, giving rise to a variety of moiré excitons with distinct electron-hole configurations \cite{ref20,ref21,ref22,ref23,ref24,ref25}. In a Chern ferromagnet, however, exciton formation takes place on a valley-polarized, time-reversal-symmetry-broken ground state. This spontaneous valley-polarized order breaks the equivalence of the two polarization-selective optical channels, which differ fundamentally from those of conventional moiré excitons \cite{ref21,ref26}. The resulting moiré Chern-ferromagnet excitons (CFEs) encode the interplay between the correlated electronic order established by \emph{e}-\emph{e} interactions and the bound electron-hole configurations reconstructed by \emph{e}-\emph{h} interactions, as illustrated in Fig.~\ref{fig:1}. Very recently, experiments have demonstrated that moiré Chern states can be probed and manipulated through excitonic optical resonances \cite{ref27,ref28,ref29}, highlighting the central role of excitonic channels in the optical response and control of Chern ferromagnets. However, despite these experimental advances, the microscopic nature of CFEs, which is essential for understanding and controlling light-matter interactions in moiré Chern ferromagnets, remains largely underexplored.

\begin{figure*}[t]
\centering
\includegraphics[width=\textwidth]{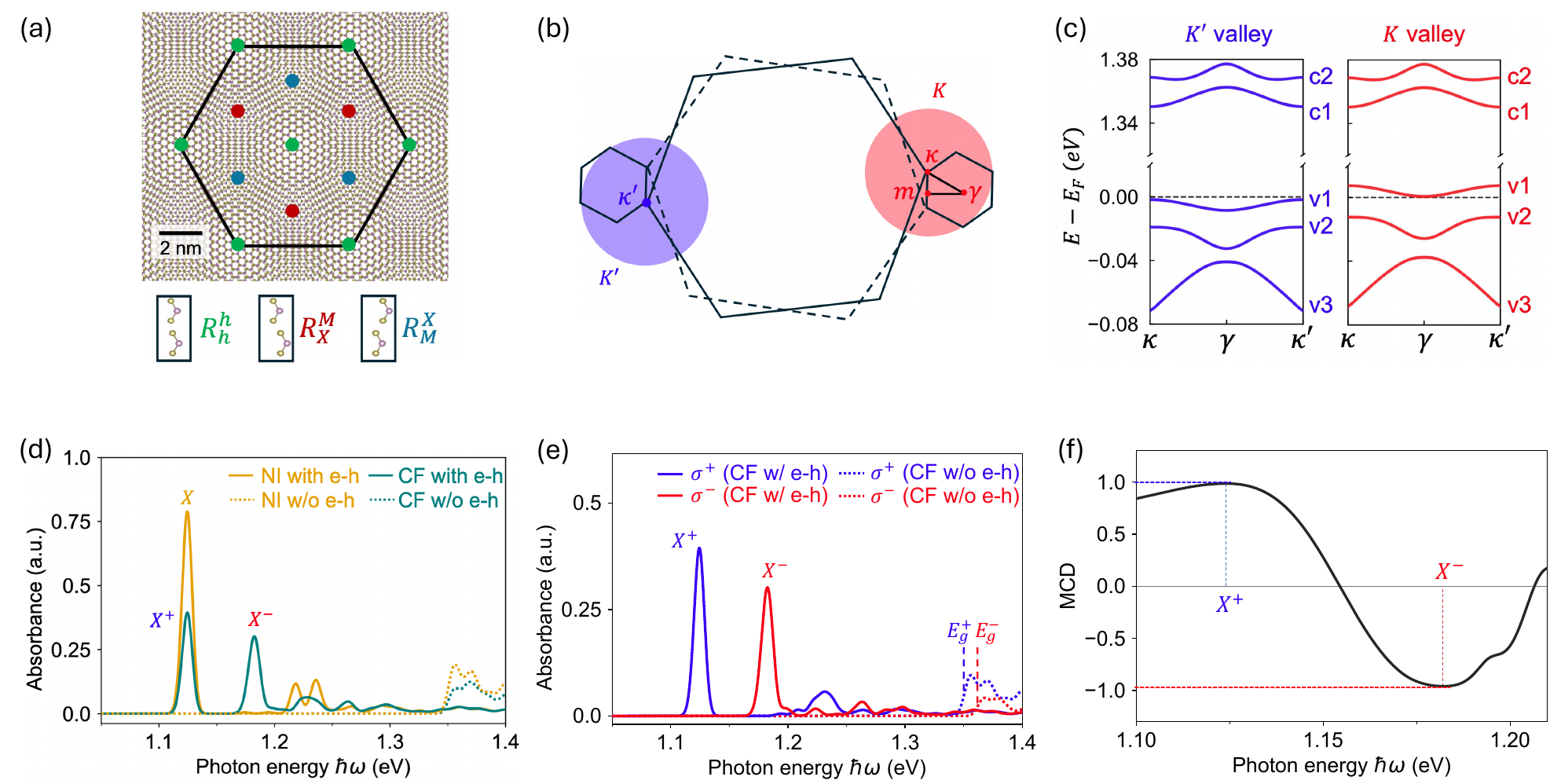}
\caption{Electronic structure and optical response of the CFEs. (a) R-type twisted bilayer MoTe\textsubscript{2} at a twist angle of  \({3.89}^{\circ}\), with a moiré period of 5.2 nm and the three high-symmetry local stackings \(R_{h}^{h}\), \(R_{X}^{M}\), and \(R_{M}^{X}\). (b) Monolayer and moiré Brillouin zones (BZ). K and K' denote the two valleys of the monolayer BZ, while \(\kappa\), \(\kappa'\), \(m\), and \(\gamma\) denote high-symmetry points of the moiré BZ. (c) Valley-resolved quasiparticle band structure of the Chern ferromagnets. v$i$ and c$i$ label the valence and conduction moiré bands, respectively. (d) Linearly polarized optical absorbance of the undoped normal insulators (NI, \(\nu_{h} = 0\)) and Chern ferromagnet (CF, \(\nu_{h} = 1\)). (e) Circularly polarized optical absorbance of the Chern ferromagnet. The two CFE resonances \(X^{+}\) and \(X^{-}\), together with the corresponding quasiparticle transition energies (band gaps)  \(E_{g}^{+}\) and \(E_{g}^{-}\), are indicated. In (d) and (e), solid (dotted) lines denote calculations with (without) \emph{e}-\emph{h} interactions. (f) Corresponding magnetic circular dichroism (MCD) of the CFEs, defined as \((\text{Abs}_{\sigma^{+}} - \text{Abs}_{\sigma^{-}})/(\text{Abs}_{\sigma^{+}} + \text{Abs}_{\sigma^{-}}),\ \)where \(\text{Abs}_{\sigma^{\pm}}\) is the absorbance for \(\sigma^{\pm}\)-polarized light.}
\label{fig:2}
\end{figure*}

In this work, we investigate moiré Chern-ferromagnet excitons (CFEs) in twisted bilayer MoTe\textsubscript{2} at a twist angle of \({3.89}^{\circ}\) with the hole filling of \(\nu_{h} = 1\). We develop a Gutzwiller plus Bethe-Salpeter equation (Gutzwiller-BSE) framework that combines the strong \emph{e}-\emph{e} correlations responsible for the valley-polarized Chern-ferromagnetic ground state with the strong \emph{e}-\emph{h} interactions governing the excitonic states. We identify two prominent near-infrared CFEs, X\textsuperscript{+} and X\textsuperscript{-}, with opposite circular-polarization selectivity. While the lower-energy X\textsuperscript{+} remains closely connected to the exciton of the undoped system, a previously underexplored CFE resonance X\textsuperscript{-} emerges and has no distinct counterpart in the undoped optical spectrum. Strong \emph{e}-\emph{h} interactions enhance the CFE energy separation and, more importantly, strongly reconstruct the internal momentum-space structure of X\textsuperscript{-} beyond quasiparticle transition picture. The reconstructed X\textsuperscript{-} state further exhibits enhanced interlayer magnitude, giving rise to a giant quadratic Stark response with a coefficient of 8.4 eV nm\textsuperscript{2} V\textsuperscript{-2}. These results reveal a nontrivial interplay between \emph{e}-\emph{e} correlations and \emph{e}-\emph{h} interactions in the formation of CFEs: electronic correlations establish inequivalent optical channels in the Chern ferromagnet, and strong \emph{e}-\emph{h} interactions subsequently reshape them into excitonic states that encode the underlying correlated order in their optical and electrical responses.

We consider R-type twisted MoTe\textsubscript{2} at twist angle \({3.89}^{\circ}\), with a moiré period of 5.2 nm [Figs.~\ref{fig:2}(a) and \ref{fig:2}(b)]. At \(\nu_{h} = 1\), the noninteracting state is valley symmetric and metallic, whereas strong \emph{e}-\emph{e} interactions spontaneously break time-reversal symmetry and stabilize a spin- and valley-polarized Chern-ferromagnetic insulating state. To determine this correlated ground state before addressing its excitonic properties, we construct an effective interacting Hamiltonian \({H = H}_{0} + H_{\text{int}}\) in the subspace of three symmetry-adapted Wannier orbitals spanning the topmost valence moiré bands \cite{ref30,ref31}. Here, \(H_{0}\) is the noninteracting Hamiltonian with parameters derived from the continuum electronic structure fitted to large-scale first-principles calculations. For the \emph{e}-\emph{e} interaction, we retain the intraorbital onsite Hubbard term,

\begin{figure*}[t]
\centering
\includegraphics[width=\textwidth]{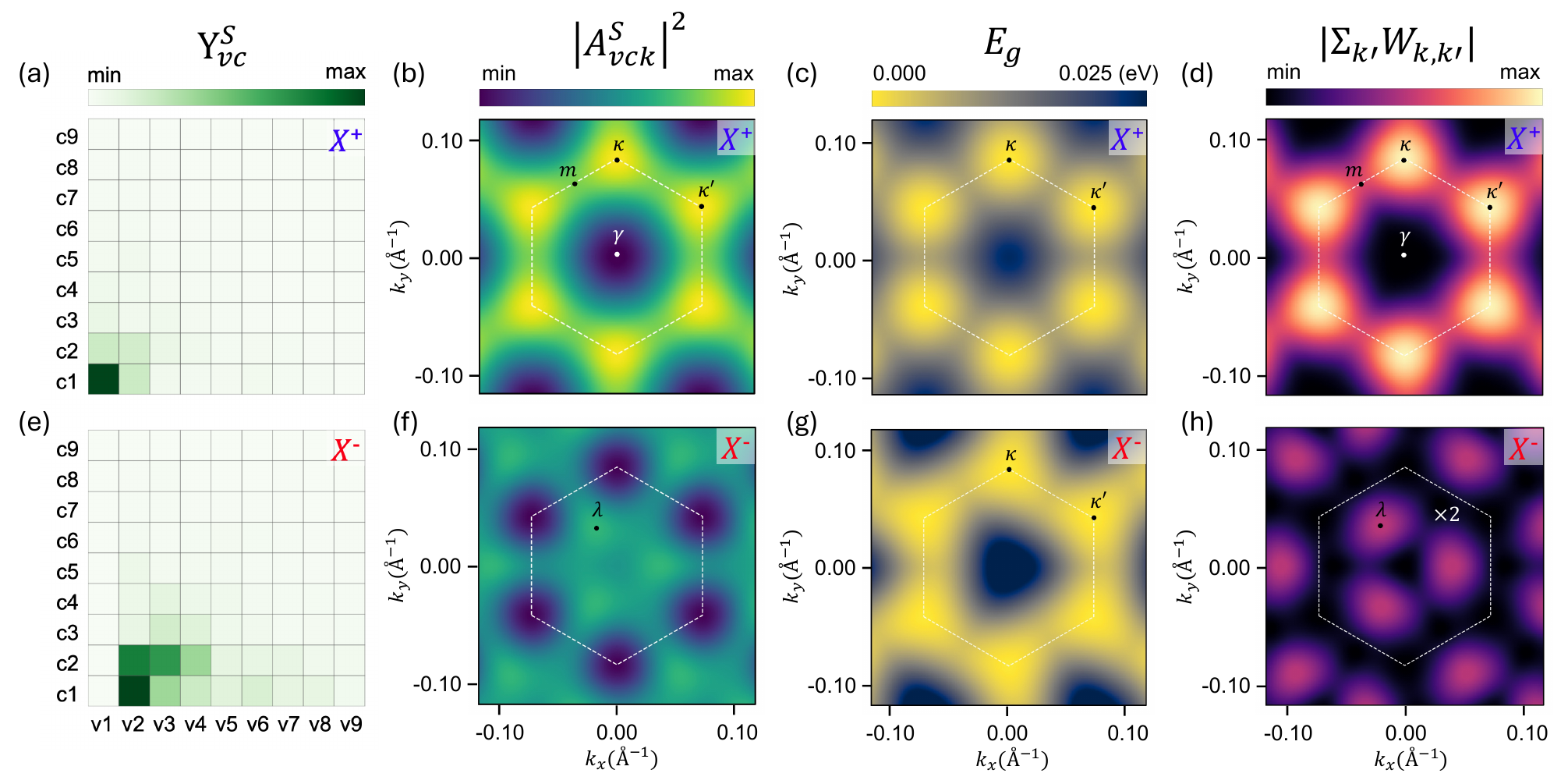}
\caption{Microscopic structure of the CFEs. \textbf{(a)}-\textbf{(d)} and \textbf{(e)}-\textbf{(h)} correspond to \(X^{+}\) and \(X^{-}\), respectively. \textbf{(a}\textbf{)},\textbf{ (}\textbf{e)} Band-pair decomposition of the CFEs, quantified by  \(Y_{\text{vc}}^{S} = \sum_{k}{\left| A_{\text{vc}}^{S}(k) \right|^{2},\ }\) identifying v1-c1 and v2-c1 as the dominant transition channels of X\textsuperscript{+} and X\textsuperscript{-}, respectively. Panels (b)-(d) and (f)-(h) characterize these dominant channels. \textbf{(b),(f) }Momentum-space distributions of CFEs, \(\left| A_{\text{vck}}^{S} \right|^{2}.\)  \textbf{(c),(g)} Quasiparticle transition energy (band gap) landscape, \({E_{g}\left( k \right) = E}_{c}\left( k \right) - E_{v}\left( k \right)\). \textbf{(d),(h)} Momentum-space projections of the screened direct electron-hole interaction kernel, \(\left| \sum_{k^{'}}W_{k,k'} \right|\). Panel \textbf{(h)} is multiplied by two for clarity. Dashed hexagons outline the moiré Brillouin zone (mBZ); \(\gamma\), \(m\), \(\kappa\) denote high-symmetry points, while \(\lambda\) marks the maximum-weight momentum of X\textsuperscript{-} in (f) and the corresponding momentum in (h).}
\label{fig:3}
\end{figure*}

\begin{equation}
H_{\text{int}} = \ U\sum_{i}{\sum_{\alpha = 1}^{3}{\widehat{n}}_{i\alpha \uparrow}}{\widehat{n}}_{i\alpha \downarrow},
\label{eq:1}
\end{equation}
where \({\widehat{n}}_{i\alpha \sigma } = {\widehat{c}}_{i\alpha \sigma }^{\dagger}{\widehat{c}}_{i\alpha \sigma }\) is the number operator for electrons with spin-valley flavor \(\sigma \ \)in orbital \(\alpha\) of the moiré cell \(i\), and \(\text{U\ }\)denote the intra-orbital onsite Coulomb repulsion. The three Wannier orbitals are centered at the high-symmetry stacking regions, \(R_{h}^{h}\), \(R_{X}^{M}\), and \(R_{M}^{X}\), shown in Fig.~\ref{fig:2}(a) and possess distinct layer character [Fig. S2 in the Supplemental Material]. We solve the interacting Hamiltonian using the Gutzwiller variational method \cite{ref32,ref33,ref34,ref35,ref36,ref37}. Within this approach, local many-body configurations are variationally reweighted by a Gutzwiller projector, yielding a time-reversal-symmetry-broken, moiré Chern-ferromagnetic ground state at \(\nu_{h} = 1\). The resulting correlation-renormalized quasiparticle states are subsequently used in the BSE calculation of the excitonic states. Details of the continuum model, Wannier construction, interaction parameters, and Gutzwiller methods are provided in the Supplemental Material.

Figure~\ref{fig:2}(c) shows the calculated valley-resolved quasiparticle band structure of the moiré Chern ferromagnet. The dominant effect of the \emph{e}-\emph{e} interaction is a valley-dependent energy renormalization of the topmost three valence moiré bands, while their overall dispersions remain largely unchanged and the band Chern topologies are preserved. Most importantly, at \(\nu_{h} = 1\), the topmost K-valley valence band v1 becomes empty of electrons (hole occupied) and is shifted above the Fermi level, whereas its K'-valley counterpart remains nearly unchanged [Fig.~\ref{fig:2}(c)]. The resulting time-reversal-symmetry-broken state is fully spin and valley polarized, with the two valleys providing inequivalent electronic configurations for optical excitation. This correlation-induced valley asymmetry produces an energy difference of approximately 10 meV between the corresponding free electron-hole optical transitions (v1 to c1 in the K' valley and v2 to v1 in the K valley), which is strongly amplified by excitonic effects as shown below.

With the Chern-ferromagnetic ground state established, we next determine its excitonic states by solving the BSE \cite{ref38,ref39}:

\begin{equation}
\begin{aligned}
&\left( E_{\text{ck}} - E_{\text{vk}} \right)A_{\text{vck}}^{S} \\ &\quad + \sum_{v^{'}c^{'}k^{'}}{K_{vck,v^{'}c^{'}k^{'}}\ A_{v^{'}c^{'}k^{'}}^{S}} = \Omega_{S}A_{\text{vck}}^{S},
\end{aligned}
\label{eq:2}
\end{equation}
where \(E_{\text{ck}}\) and \(E_{\text{vk}}\) are the quasiparticle conduction and valence band energies, \(\Omega_{S}\) is the exciton energy for exciton \(S\), and \(A_{\text{vck}}^{S}\) is the exciton wavefunction in the electron-hole basis. The interaction kernel \(K = V - W\) contains the bare exchange interaction \(V\) and the screened direct Coulomb attraction \(W\) (see Supplemental Material for details). The computed optical spectra are shown in Figs.~\ref{fig:2}(d) and \ref{fig:2}(e). Figure~\ref{fig:2}(d) compares the linearly polarized absorbance of the Chern ferromagnet at \(\nu_{h} = 1\ \)with that of the undoped normal insulator (NI) at \(\nu_{h} = 0\). In the NI, time-reversal symmetry preserves the valley degeneracy, giving rise to a single dominant excitonic resonance X at approximately 1.12 eV, with equivalent contributions from the two valleys. In the Chern ferromagnet, by contrast, the broken valley symmetry gives rise to two well-resolved CFE resonances, X\textsuperscript{+} and X\textsuperscript{-}. Circularly polarized absorbance [Fig.~\ref{fig:2}(e)] further reveals their opposite helicity selectivity: X\textsuperscript{+} couples selectively to \(\sigma^{+}\) light, whereas X\textsuperscript{-} couples to \(\sigma^{-}\) light. This strong circular-polarization selectivity gives rise to nearly perfect magnetic circular dichroism (MCD), with opposite signs at the two CFE resonances [Fig.~\ref{fig:2}(f)]. Analysis of the exciton wavefunctions shows that X\textsuperscript{+} and X\textsuperscript{-} predominantly originate from the K' and K valleys, respectively. The opposite valley origins of X\textsuperscript{+} and X\textsuperscript{-} give rise to exciton Berry curvatures \cite{ref40} of opposite signs, offering a promising means of distinguishing and spatially separating the two CFEs through the exciton Hall effect (see Fig. S3 in the Supplemental Material for details).

The lower-energy X\textsuperscript{+} resonance remains close in energy to the exciton X of the undoped system and carries approximately half of its linearly polarized absorbance, consistent with the nearly unchanged K'-valley electronic structure [Figs.~\ref{fig:1} and \ref{fig:2}(c)]. In contrast, a previously underexplored CFE resonance X\textsuperscript{-} emerges from the K valley of the Chern-ferromagnetic state, which is well-separated from X\textsuperscript{+} and has no distinct counterpart in the undoped optical spectrum. Most strikingly, the two CFE resonances are separated by about 60 meV and exhibit strongly contrasting helicity-resolved optical responses [Fig.~\ref{fig:2}(e)]. However, in the absence of \emph{e}-\emph{h} interactions, the corresponding quasiparticle transition energies \(E_{g}^{+}\) and \(E_{g}^{-}\) differ by only 10 meV [dashed curves in Fig.~\ref{fig:2}(e)]. Strong \emph{e}-\emph{h} interactions therefore enhance the energy separation by nearly a factor of 6, rendering the two valley-selective resonances readily distinguishable in optical spectroscopy. More fundamentally, the pronounced enhancement demonstrates that the optical response of the Chern ferromagnet cannot be understood from quasiparticle transitions alone, and thus highlights the crucial role of strong \emph{e}-\emph{h} interactions in the formation of CFEs.

To uncover the microscopic origin of the enhanced CFE splitting and their contrasting optical character, we examine the internal structure of the two CFEs in Fig.~\ref{fig:3}. Figures~\ref{fig:3}(a) and \ref{fig:3}(e) show their band-pair decomposition, \(Y_{\text{vc}}^{S} = \sum_{k}{\left| A_{\text{vc}}^{S}(k) \right|^{2},\ }\) using the same c-v band labels for the two valleys as in Fig.~\ref{fig:2}(c). X\textsuperscript{+} is dominated by the v1-c1 transition channel in the K' valley, whereas X\textsuperscript{-} is dominated by the v2-c1 channel in the K valley. At the K valley, our BSE calculations also reveal a separate class of very low-energy intra-valence excitons ($\sim$ tens of meV) arising predominantly from v2-v1 transitions. These excitations lie in the terahertz energy range ($\sim$ 10 THz) \cite{ref41,ref42}, well separated from the near-infrared CFEs considered here, and remain essentially unmixed with them. Throughout this paper, we focus on the near-infrared CFEs, which are also the principal optical excitations of interest in recent experiments \cite{ref27,ref28,ref29}.

\begin{figure}[tbp]
\centering
\includegraphics[width=\columnwidth]{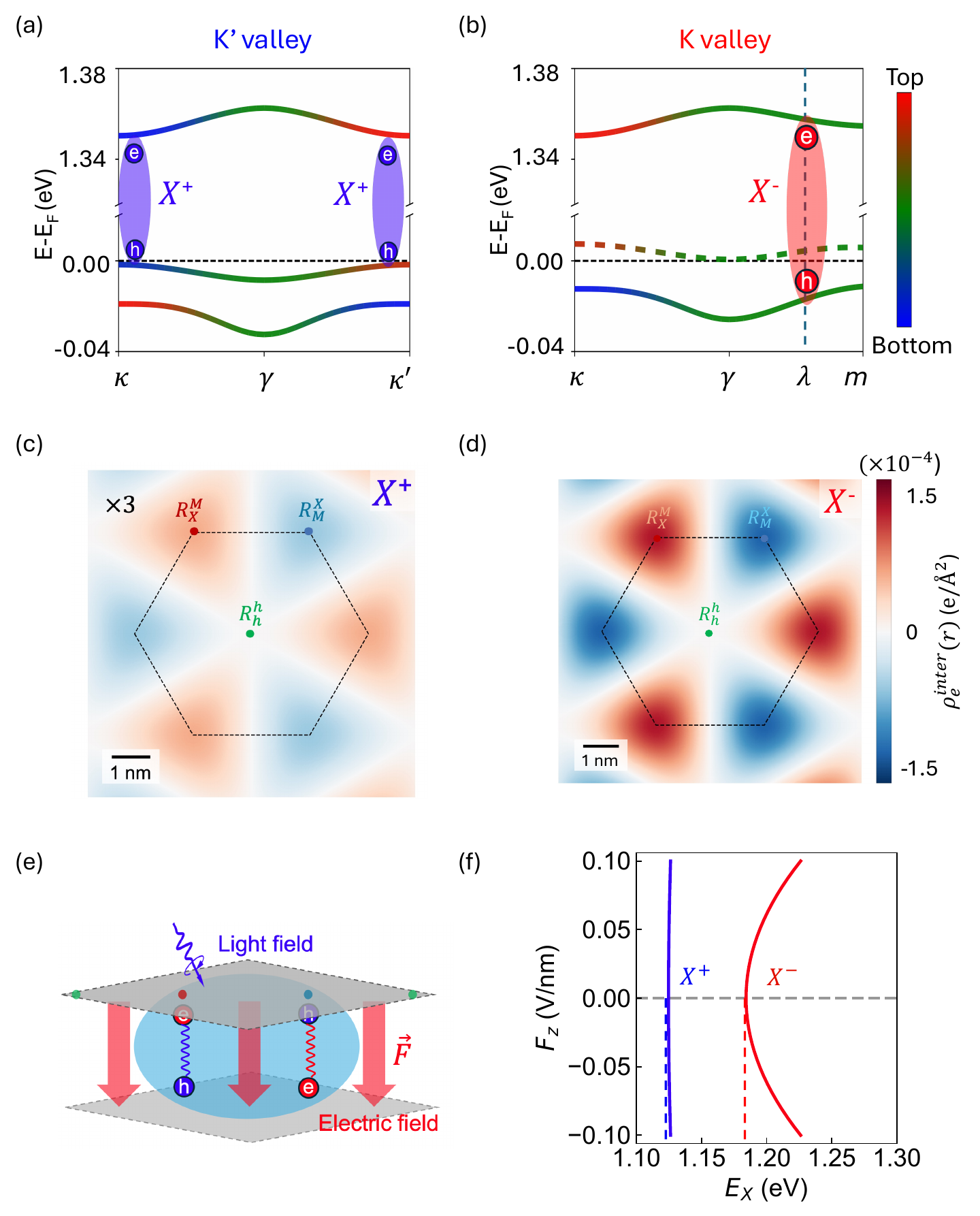}
\caption{Interlayer character and quadratic Stark response of the CFEs. (a),(b) Layer polarization of the quasiparticle states associated with (a) \(X^{+}\) in the K' valley and (b) \(X^{-}\) in the K valley. The \(e\) and \(h\) markers indicate the dominant electron-hole transition channels. (c),(d) Real-space distributions of the interlayer electron density \(\rho_{e}^{\text{inter}}(r)\) for (c) \(X^{+}\ \) and (d) \(X^{-}\). Positive (negative) values correspond to electrons in the bottom (top) layer, with the hole position integrated over the opposite layer. Panel (c) is multiplied by a factor of three for clarity. (e) Schematic illustration of the excitonic quadratic Stark effect setup under a uniform out-of-plane electric field \(F_{z}\). (f) CFE energies as a function of \(F_{z}\ \)for  \(X^{+}\ \)(blue) and \(X^{-}\ \) (red), showing a strongly enhanced quadratic Stark response for \(X^{-}.\) Dashed vertical lines mark the zero-field exciton energies.}
\label{fig:4}
\end{figure}

The two CFEs exhibit markedly different momentum-space structures. For their respective dominant transition channels, X\textsuperscript{+} is concentrated around \(\kappa\) (and \(\kappa'\)), whereas the weight of X\textsuperscript{-} is strongly suppressed near \(\kappa\) (and \(\kappa'\)) and reaches its maximum at \(\lambda \ \) [Figs.~\ref{fig:3}(b) and \ref{fig:3}(f)]. To identify the origin of this contrast, we separately investigate the two main contributions to the BSE Hamiltonian [Eq.~\eqref{eq:2}]: the quasiparticle transition energy \(E_{g}\left( k \right),\ \)and the \emph{e}-\emph{h} interaction kernel, which is dominated by the screened direct Coulomb attraction \(W\). For both X\textsuperscript{-} and X\textsuperscript{+}, the \(E_{g}\left( k \right)\) landscapes exhibit a band-gap minimum at \(\kappa\) (and \(\kappa'\)) [Figs.~\ref{fig:3}(c) and \ref{fig:3}(g)]. Therefore, quasiparticle transition energies alone favor the \(\kappa\) (and \(\kappa'\)) region for both CFEs and cannot explain why X\textsuperscript{-} reaches its maximum at \(\lambda\). In contrast, the momentum-space projections of \(W\) show qualitatively different structures: the interaction is strongest around \(\kappa\) (and \(\kappa'\)) for X\textsuperscript{+}, but around \(\lambda\) for X\textsuperscript{-} [Figs.~\ref{fig:3}(d) and \ref{fig:3}(h)], closely tracking their corresponding excitonic weight distributions in Figs.~\ref{fig:3}(b) and \ref{fig:3}(f). This correspondence shows that the unconventional momentum-space structure of X\textsuperscript{-} is driven by strong \emph{e}-\emph{h} interactions, rather than inherited from the quasiparticle band transition. Moreover, comparison of Figs.~\ref{fig:3}(d) and \ref{fig:3}(h) shows that the screened direct Coulomb attraction W is substantially stronger for X\textsuperscript{+} than for X\textsuperscript{-}, giving rise to different binding energies that contribute to the large CFE energy separation discussed above. These results reveal a nontrivial interplay between \emph{e}-\emph{e} and \emph{e}-\emph{h} interactions underlying the formation of CFEs. The valley-polarized Chern-ferromagnetic ground state established by \emph{e}-\emph{e} correlations provides inequivalent optical channels, on top of which strong \emph{e}-\emph{h} attraction amplifies the excitonic splitting and reconstructs the internal exciton wavefunctions.

The excitonic-interaction-reconstructed microscopic character of emerging X\textsuperscript{-} is further manifested in the real-space properties. As shown by the layer-resolved quasiparticle states in Figs.~\ref{fig:4}(a) and \ref{fig:4}(b), the v1 and c1 states contributing to X\textsuperscript{+} near \(\kappa\) (and \(\kappa'\)) reside predominantly in the same layer, whereas the v2 and c1 states contributing to X\textsuperscript{-} near \(\lambda\) exhibit pronounced interlayer hybridization. This difference suggests distinct interlayer electron-hole characteristics for the two CFEs. To characterize their real-space distributions, we define a signed interlayer electron density, $\rho_{e}^{\text{inter}}(\mathbf r_e)
={}\int\bigl[|\chi_{\text{bt}}(\mathbf r_e,\mathbf r_h)|^2\\
\quad-|\chi_{\text{tb}}(\mathbf r_e,\mathbf r_h)|^2\bigr]d\mathbf r_h$, where \(\chi_{\text{bt}}\) (\(\chi_{\text{tb}})\) denotes the exciton wavefunction with the electron in the bottom (top) layer and the hole in the top (bottom) layer. The magnitude reflects the strength of the local interlayer excitonic component, while its sign distinguishes the two opposite interlayer electron-hole orientations. As shown in Figs.~\ref{fig:4}(c) and \ref{fig:4}(d), both CFEs exhibit interlayer electron densities concentrated around the \(R_{X}^{M}\) and \(R_{M}^{X}\) stacking regions, with alternating signs between these regions. Notably, the interlayer density amplitude is approximately 7 times larger for X\textsuperscript{-} than for X\textsuperscript{+}, revealing the substantially enhanced interlayer character of the emergent X\textsuperscript{-} state.

These distinct interlayer characteristics are also reflected in the response of the CFEs to an out-of-plane electric field, as illustrated in Fig.~\ref{fig:4}(e). The joint symmetry of \(C_{2y}\) and time reversal operator \(\mathcal{T}\) yields a vanishing net perpendicular electric dipole for both CFEs, and thus suppressing their linear Stark response. However, the coherent interlayer components of both states give rise to a quadratic response to a static out-of-plane electric field \(F_{z}\),

\begin{equation}
{\Delta E}_{S}(F_{z}) = \frac{1}{2}\kappa _{S}F_{z}^{2},
\label{eq:3}
\end{equation}
where \({\Delta E}_{S}\) is the Stark energy shift and \(\kappa _{S}\ \)is the quadratic Stark coefficient, evaluated from second-order perturbation theory as detailed in the Supplemental Material. Figure~\ref{fig:4}(f) shows the magnitude of \(\kappa _{X^{-}}\) exceeds \(\kappa _{X^{+}}\) by more than a factor of 20 and reaches 8.4 eV nm\textsuperscript{2} V\textsuperscript{-2}, consistent with the enhanced interlayer character of X\textsuperscript{-}. Together with their opposite circular-polarization selectivity, these sharply contrasting quadratic Stark responses leave complementary spectroscopic fingerprints in helicity- and field-resolved optical spectra, providing experimentally accessible signatures for identifying and manipulating the emerging CFEs. We also note that the magnitude of \(\kappa _{X^{-}}\) is comparable to the giant quadratic Stark responses recently observed for quadrupolar excitons in other TMD bilayers \cite{ref43,ref44,ref45,ref46}, placing the predicted response well within an experimentally accessible regime.

In summary, we have established a microscopic picture of Chern-ferromagnet excitons (CFEs) in twisted bilayer MoTe\textsubscript{2} by treating the correlated electronic ground state and excitonic interactions within the developed Gutzwiller-BSE framework. We identify two near-infrared CFEs with opposite circular-polarization selectivity whose energy separation reaches about 60 meV, despite a difference of only about 10 meV between the corresponding quasiparticle transitions. The higher-energy X\textsuperscript{-} resonance is particularly striking: it has no distinct counterpart in the undoped optical spectrum, and its internal momentum-space structure is strongly reconstructed by \emph{e}-\emph{h} interactions beyond the quasiparticle-transition picture. This reconstructed state also exhibits substantially enhanced interlayer character, giving rise to a quadratic Stark response more than 20 times larger than that of X\textsuperscript{+}. These complementary helicity- and field-dependent responses provide experimentally accessible spectroscopic fingerprints of CFEs and establish excitonic spectroscopy as a microscopic probe of interaction-driven Chern ferromagnetism in moiré materials.

\begin{acknowledgments}

We thank Xiangbin Cai, Kam Tuen Law, Fengcheng Wu, Hong Guo, Wang Yao, Tianyu Qiao, Wangqian Miao, Hanqi Pi and Fazhi Yang for useful discussions. This work is supported by the start-up fund and the Direct Grant (No. 4053726) of the Chinese University of Hong Kong, the National Natural Science Foundation of China (No. 12504222), and the 1+1+1 CUHK-CUHK(SZ)-GDSTC Joint Collaboration Fund (GRDP2026-018), and the State Key Laboratory of Quantum Information Technologies and Materials.

\end{acknowledgments}


\bibliography{references}

\end{document}